\documentclass[pre]{revtex4}
\usepackage{graphicx}
\usepackage{latexsym}
\usepackage{amsmath}
\usepackage{amssymb}
\usepackage{amsfonts}
\usepackage{amsthm}

\begin{document}
\title{Bending of solitons in weak and slowly varying inhomogeneous plasma}
\author{ Abhik Mukherjee
\footnote{abhik.mukherjee@saha.ac.in}
, M.S. Janaki
\footnote{ms.janaki@saha.ac.in}
, Anjan Kundu
\footnote{anjan.kundu@saha.ac.in}
}
\affiliation{Saha Institute of Nuclear Physics\\
 Kolkata, INDIA}

\begin{abstract} 
Bending of  solitons in two dimensional plane is presented in the presence of
weak and slowly varying inhomogeneous ion density for the propagation of ion acoustic soliton
in unmagnetized cold plasma with isothermal electrons. Using
reductive perturbation technique, a modified Kadomtsev- Petviashvili  equation is obtained
with a chosen unperturbed ion density profile. Exact solution of the equation shows
 that the phase of the solitary wave gets
modified by a function related to the unperturbed inhomogeneous ion density causing
 the soliton to bend in the two dimensional plane, whereas the amplitude of the soliton
remaining constant.

\end{abstract}
\maketitle

\smallskip


\section{Introduction}	Extensive investigations of ion acoustic solitons in plasma were started,
since Washimi and Taniuti \cite{Washimi} showed that ion acoustic waves in a weakly nonlinear dispersive plasma
could be described by Korteweg-de Vries (KdV) equation. Since then plasma physics community
has been actively involved in nonlinear phenomena related structures such as solitons, shocks, phase-space holes
etc \cite{Chen}. In a homogeneous plasma, an ion acoustic soliton travels without change in shape, amplitude
and speed \cite{Chen,Farah}. But in actual experimental conditions, we encounter inhomogeneities
in plasma at the edges or boundaries of the system or in the presence of density gradient.
The propagation of ion acoustic KdV solitons in an inhomogeneous plasma was first considered by Nishikawa and Kaw
\cite{Kaw} who presented a WKB solution when its spatial width is very small as compared to 
density gradient scale length. Gell and Gomberoff \cite{Gell} reconsidered the situation
and showed that amplitude, velocity and width of the soliton are proportional to the fractional powers of ion
density which was verified experimentally by John and Saxena\cite{JohnSaxena} and modified by Rao and Verma
\cite{RaoVerma} by taking into account ion drift velocity, but allowing terms proportional to the 
stretched variable $\xi$ in their first order equations. These inconsistencies were later removed by Kuehl
and Imen \cite{KuehlImen} and their results are found to be in good agreement with those of Chang
et.al\cite{Chang}.
One of the most important features of ion acoustic soliton is its reflection by plasma inhomogeneity.
This phenomenon was first observed experimentally by Dahiya et.al\cite{Dahiya2} from the sheath around a negatively
biased grid, where the density gradient is high. Popa and Oertl found reflection of 
ion acoustic soliton from a bipolar potential wall structure\cite{Popa}, Nishida \cite{Nishida} and Imen - Kuehl
\cite{ImenKuehl} found from a finite plane boundary, Nagasawa \cite{Nagasawa} found from a metallic mesh
electrode showing nonlinear Snell's law and Yi and Cooney et.al found from a sheath in a negative ion plasma
\cite{Seungjun, Cooney}. Kuehl investigated theoretically the reflection of ion acoustic soliton,
and showed that a shelf develops behind the soliton and the reflected wave is small compared with
both trailing shelf and soliton amplitude decrease due to energy transfer to the shelf\cite{Kuehl, KoKuehl,KoKuehl2}.
Then after, many authors took the problem of soliton propagation in inhomogeneous plasma in different 
physical situations like plasma with finite ion temperature\cite{SinghDahiya,Singhdahiya2},
with negative ions \cite{Malik1,Malik6,Chauhan,SinghMalik}, with dust\cite{Xiao} and trapped electrons
\cite{Kumar}, in magnetic field \cite{Malik3,Mahmood}, with non isothermal electrons \cite{Malik4},
with ionization \cite{Malik5,Malik8}, with electron inertia contribution \cite{Malik7} and also in
other contexts \cite{Duan-Zhao,IbrahimKuehl,Cooney}.

These discussed cases are all (1+1) dimensional, but in practical circumstances the waves observed in laboratory
and space are certainly not bounded in one dimension. Nevertheless, the two dimensional propagation 
of ion acoustic waves in inhomogeneous plasma has received much less attention. Zakharov- Kuznetsov
(ZK) equation,  which is the more isotropic 2 dimensional generalization of KdV equation, was obtained in modified
form in magnetized dusty inhomogeneous plasma with non-extensive electrons \cite{ZK2}, with
dust charge fluctuation\cite{ZK3}, with quantum effects \cite{ZK4}, with non thermal ions and dust charge
variation\cite{ZK1} and in other situations. But if weak transverse propagation is considered
then the possible 2 dimensional generalization of KdV model is Kadomtsev- Petviashvili (KP) equation which was first
derived in the context of plasma\cite{KP}. Malik et.al derived KP equation in modified form 
 in inhomogeneous plasma with finite temperature drifting ions \cite{KP2}
and solved it for constant density gradient. Later in quantum inhomogeneous plasma a modified KP equation
was also obtained \cite{KP3} and line soliton solutions were presented. Along
with reflection and transmission of line solitons in inhomogeneous plasma, its bending in two dimensional plane
is also a possible relevant phenomenon which was not  explored in literature considered earlier as well as in
 \cite{YANG,ZhangXue}  in two dimensions as far as  our knowledge goes.

In this brief communication, we have taken up this problem by considering
ion acoustic soliton propagation in unmagnetized, cold plasma
with hot isothermal electrons. Using reductive perturbation technique,
a modified form of KP equation is obtained for weak transverse propagation
and weak and slowly varying inhomogeneous ion number density . Exact solitary wave solutions were presented
showing the bending of ion acoustic solitons in two dimensional
plane. The soliton is modified in phase which is controlled by a function related to
equilibrium ion number density,  causing soliton bending in two dimensional plane, whereas the amplitude
remains constant. 
The paper is organized as follows. The derivation of the corresponding evolution equation in 2
 dimensional plane
for a weak and slowly varying inhomogeneous plasma is given in Sec-II. Sec-III deals with the phase
modulated solitary wave solutions showing bending in 2 dimensional
 plane. Conclusions and remarks are contained in Sec-IV.

\section{
Derivation of two dimensional evolution equation for an ion acoustic wave propagating in a 
weak and slowly varying inhomogeneous plasma}
We consider a two dimensional, collisionless, unmagnetized, weak and slowly varying spatially 
inhomogeneous plasma consisting of hot isothermal electrons and cold ions ($T_i = 0$).
 The plasma is weakly inhomogeneous with a slow  variation of the equilibrium ion density
along one spatial direction.  The ion continuity and momentum 
 equations together with Poisson's equation and the electron Boltzmann distribution can be written
in the dimensionless form as

\begin{equation}
\frac{\partial n}{\partial t}+\overrightarrow{\bigtriangledown}\cdot(n \overrightarrow u ) =0
,\ \
\frac{\partial \overrightarrow u}{\partial t}+
(\overrightarrow u\cdot\overrightarrow{\bigtriangledown})\overrightarrow u +
\overrightarrow \bigtriangledown \phi  =0, \ \
\bigtriangledown^2 \phi =  n_e - n, 
n_e = \exp(\phi) \label{basic}
\end{equation} 

 In equation (\ref{basic}), $u\equiv (u_x,u_y)$ is the ion fluid velocity
normalized by the ion acoustic speed $c_s= \sqrt{\frac{T_e}{m_i}}$, n and $n_e$
are ion and electron number densities respectively normalized by unperturbed ion number density $\tilde{n_0}$
 at an arbitrary reference point in plasma which we chose to be $x=0$,
$\phi$ is the electrostatic potential normalized by $\frac{T_e}{e}$ where $T_e, m_i, e$
are electron temperature, ion mass and electronic charge respectively. All the spatial co-ordinates
$x,y$ are normalized by the Debye length $\lambda_D=\sqrt{\frac{\epsilon_0 T_e}{\tilde{n_0} e^2}}$ at 
$x=0$
and time by inverse of the ion plasma frequency $\omega_{pi}= \sqrt{\frac{\tilde{n_0} e^2}{\epsilon_0 m_i}}$
at $x=0$,
where $\epsilon_0$ is the permettivity of free space. We have assumed that the equilibrium
electron and ion number densities are equal at $x=0$ (quasi-neutrality) and that the zero reference
of the equilibrium potential is at $x=0$.
In the above equations, the ions are assumed to be cold and on the slow ion time scale, the
 electrons are assumed to be in local thermodynamic equilibrium.
When the electron inertia is neglected, the electrons can be considered to follow a Boltzmann distribution.

Under these assumptions, in the absence of any equilibrium drift, the ion acoustic waves follow the 
dispersion relation given by
\begin{equation}
 \omega = k c_s,
\end{equation}
where $\omega, k, c_s$ are angular frequency, wave vector and ion acoustic speed respectively.

In order to study the ion-acoustic wave propagation and its two dimensional evolution as a solitary wave
in weak and slowly varying inhomogeneous plasma, we consider the following appropriate stretched co-ordinates
 \begin{eqnarray}
\xi= \epsilon^{\frac{1}{2}} (x- M t), \ \
\lambda= \epsilon y, \ \
\eta= \epsilon^{\frac{3}{2}} x,
\label{scaling}\end{eqnarray} 
where $M$ is a constant  which is the phase velocity of the wave normalized
by the ion acoustic speed and $\epsilon$ is a small expansion parameter. Generally, phase velocity 
is taken to be a function  of x in the literatures of inhomogeneous plasma, but here we have taken
$M$ to be a constant which is similar as the scaling used by Gell in \cite{Gell}. This assumption will
be shown to be consistent with the calculations for the chosen unperturbed ion number density profile.

Chang et.al, in their experimental studies of propagation of ion acoustic solitons in an inhomogeneous plasma
\cite{Chang},
created a definite ion number density profile as shown in FIG.1, in a large multi dipole plasma device.
To create local inhomogeneity in a previously homogeneous quiscent plasma, a perturbing object was inserted
far from the excitation region. 
The left portion of FIG.1, where the density variation
is slow was shown to be the host environment for studying
soliton characteristics. It was also reported that the experiment had a pronounced two dimensional 
character. 

We have followed the experimental results  of Chang et.al 
and numerical solutions obtained by Kuehl [8],[13] 
in the context of the propagation of ion acoustic wave in inhomogeneous plasma.
 The ion density plot, which was reported in their paper is reproduced here as FIG.1.

It is evident from the figure that, before introduction of the perturbing structure the density
was homogeneous (continuous curve) and the presence of the structure caused a density inhomogeneity
as shown by the broken curve. This plot is consistent with the numerical solutions
of the equilibrium density shown in [8] and [13].

Following the above stated environment for soliton propagation, we have taken the unperturbed ion
number  density
profile to be of the form $\tilde{n_0}(\eta) = 1+ \delta  f_0(\eta)$,
where $\delta$ is a small parameter, having the same features
of the left portion of FIG.1 . The inhomogeneity is weak
 as well as slowly varying along $\eta$ as shown in FIG.1, so that the plasma 
is nearly homogeneous. Our entire work is based 
on this region of weak and slowly varying inhomogeneity, showing more finer effects on the propagation of soliton.
 Experimental methods of producing such density gradients have been discussed  in earlier works\cite{JohnSaxena,
Dahiya2}.
\begin{figure}
 \includegraphics[height=5cm,width=7cm]{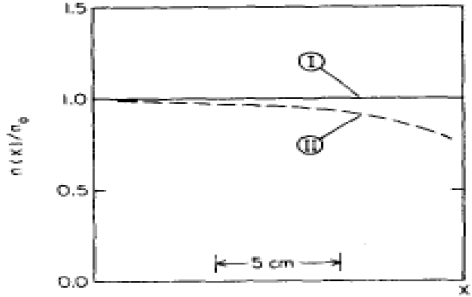}
\caption{Measured ion number density profile in the target chamber for the experiment
done by Chang et.al \cite{Chang}. Continuous line denotes the profile if the perturbing 
structure is absent and the broken line denotes if it is present at the right edge.
 Reproduced with permission from Phys.Fluids 29,294(1986).Copyright 1986 American Physical Society.}
\end{figure}

From the steady state condition of ion continuity equation we get
\begin{equation}
 \frac{\partial}{\partial \eta}[\tilde{n_0}\tilde{u_0}] = 0,
\end{equation}
where $\tilde{u_0}$ is the equilibrium ion velocity. Hence after integration $\tilde{u_0}$
can be determined as 
\begin{equation}
 \tilde{u_0} = \frac{c_1}{\tilde{n_0}} = c_1[1 - \delta f_0],
\end{equation}
where $c_1$ is an integration constant and higher order terms are neglected due to smallness.

Now from the steady state condition of the x component of momentum equation we get
\begin{equation}
 \frac{\partial}{\partial \eta}[\tilde{u_0}^2 + \tilde{\phi_0}] = 0,
\end{equation}
where $\tilde{\phi_0}$ is the equilibrium potential.
After integration $\tilde{\phi_0}$ can be determined as
\begin{equation}
 \tilde{\phi_0} = C_2 - {\tilde{u_0}}^2 ,
\end{equation}
where $C_2$ is another integration constant. Choosing $C_2 = c_1^2$, we get
 \begin{equation}
 \tilde{\phi_0} = 2 c_1^2 \delta f_0(\eta) ,
\end{equation}
where also higher order terms are neglected due to smallness.  Choosing this, 
we can also see that the steady state condition of Poisson's equation is also satisfied
for these functions of $\tilde{\phi_0}, \tilde{n_0}$ if the higher order terms are neglected 
due to smallness.

 These equilibrium quantities are obtained self consistently
from the fluid equations (\ref{basic}).
 To create equilibrium in a real 
experimental situation, external electric fields are imposed by using  appropriate biasing arrangements
 inside the plasma.  
Details of the setup are found in  \cite{JohnSaxena, Dahiya2}.  
This also gives rise to steady drift that is space dependent in presence of density gradients.

 The equilibrium electron number density is also inhomogeneous. 
 The inhomogeneity of the  equilibrium electron density
can be expressed clearly from equation (\ref{basic}), from where we get
\begin{equation}
 \tilde{n_{e0}} = e^{\tilde{\phi_0}(\eta)},
\end{equation}
hence it is also inhomogeneous.

A reductive perturbation method is carried out with $\epsilon$ as the expansion
parameter to obtain the two dimensional nonlinear
evolution equation with weak transverse propagation.  $\delta$  is a
small parameter which is controlled externally to form the equilibrium density profile. For the sake of this
work we take here $\delta$ to be $ \approx \epsilon$.

All the variables are expanded as
\begin{equation}
n= 1 + \epsilon f_0(\eta) + \epsilon n_1(\xi,\eta,\lambda) + \epsilon^2 n_2(\xi,\eta,\lambda)+...
\label{n}\end{equation}
\begin{equation}
\phi = 2 c_1^2 \epsilon f_0(\eta) + \epsilon \phi_1(\xi,\eta,\lambda) + \epsilon^2 \phi_2(\xi,\eta,\lambda)+...
\label{phi}\end{equation}
\begin{equation}
u_{x} =  c_1[1 - \epsilon f_0(\eta)] + \epsilon u_1(\xi,\eta,\lambda) + \epsilon^2 u_2(\xi,\eta,\lambda)+...
\label{ux}\end{equation}
\begin{equation}
u_{y} =   \epsilon^{\frac{3}{2}} v_1(\xi,\eta,\lambda) + \epsilon^{\frac{5}{2}} v_2(\xi,\eta,\lambda) +...
\label{uy}\end{equation}

The set of stretched quantities and the expansion 
of the physical quantities given by (\ref{scaling})and (\ref{n})-(\ref{uy}) are used
in the fluid equations (\ref{basic}) and the coefficients of different
powers of $\epsilon$  are collected and set to zero.

At the lowest order $\bf{\epsilon}$,  we get
\begin{equation}
\phi_{1} + 2 c_1^2 f_0 =  n_{1} + f_0
 \label{epsilon}\end{equation}

At  $\bf{\epsilon^{\frac{3}{2}}}$ we get,

\begin{equation}
(M - c_1) \frac{\partial n_1}{\partial \xi} = \frac{\partial u_1}{\partial \xi} ,\ \
\frac{\partial \phi_1}{\partial \xi}= (M - c_1) \frac{\partial u_1}{\partial \xi},
\label{epsilon3/2}
\end{equation}
from where we obtain $(M - c_1) n_1 = u_1$ and $(M - c_1) u_1 = \phi_1$, where it is assumed that
as $\xi \rightarrow \pm \infty$, $n_1, u_1\rightarrow 0$. Using (\ref{epsilon}) and (\ref{epsilon3/2})
we get $(M - c_1)^2 =1$  and $2 c_1^2 = 1$ which gives $c_1 = \frac{1}{\sqrt{2}}$ 
and $\phi_1= n_1 = u_1$.

Because of the presence of  drift, the equilibrium dispersion relation
in the normalized variable $M$ is given by
 $M = 1 + \frac{1}{\sqrt{2}}$.

At  $\bf{\epsilon^2;}$ we obtain,

\begin{equation}
  \frac{\partial v_{1}}{\partial \xi}= \frac{\partial \phi_{1}}{\partial \lambda},\ \
\frac{\partial^2 \phi_{1}}{\partial \xi^2} +n_2 = \phi_2 + \frac{1}{2}(f_0 + \phi_1)^2
\label{epsilon2}
 \end{equation}

Finally at  $\bf{\epsilon^{\frac{5}{2}}}$ order we get

\begin{equation}
  -\frac{\partial n_{2}}{\partial \xi}
+\frac{\partial}{\partial \xi}(-c_1 f_0 n_1 + u_1 n_0)
+ \frac{\partial u_{2}}{\partial \xi} + \frac{\partial}{\partial \xi}(u_1 n_1 ) -c_1
\frac{\partial f_{0}}{\partial \eta}
+\frac{\partial u_{1}}{\partial \eta}+\frac{\partial v_{1}}{\partial \lambda}
+ c_1 \frac{\partial f_{0}}{\partial \eta} + c_1 \frac{\partial n_{1}}{\partial \eta}=0,
\end{equation}
\begin{equation}
 -\frac{\partial u_{2}}{\partial \xi}
+\frac{\partial}{\partial \xi}(-c_1 f_0 n_1 )
+ \frac{\partial \phi_{2}}{\partial \xi} + n_1 \frac{\partial}{\partial \xi}( n_1 ) -c_1
\frac{\partial f_{0}}{\partial \eta}
+ c_1\frac{\partial u_{1}}{\partial \eta}
+ 2 c_1^2 \frac{\partial f_{0}}{\partial \eta} + \frac{\partial n_{1}}{\partial \eta}=0,
\label{epsilon5/2}
\end{equation}
combination of which  using (\ref{epsilon2}), we get the final evolution 
equation
\begin{equation}
 \frac{\partial}{\partial \xi} [(2 + 2 c_1)\frac{\partial n_{1}}{\partial \eta}
+2 n_1 \frac{\partial n_1}{\partial \xi}
+ \frac{\partial^3 n_{1}}{\partial \xi^3}] +
\frac{\partial^2}{\partial \lambda^2}(n_1)
-2 c_1 f_0\frac{\partial^2 n_{1}}{\partial \xi^2}=0,\ \
\label{kp1}
\end{equation} 
with $c_1 = \frac{1}{\sqrt{2}}$, which is nothing but Kadomtsev-Petviashvili (KP) equation with an extra term
appearing due to inhomogeneity . Here we have considered the simplest
configuration of unmagnetized plasma with cold ions and isothermal electrons, but the similar equation with 
different coefficients can be derived for more complexities like ion temperature, presence of magnetic field etc
for the chosen equilibrium ion number density profile.
Note that, in \cite{KP2} the modified KP equation was derived, considering the fact that the scale length
of the plasma inhomogeneity is much larger than the width of the soliton, and solitary wave 
solution is given for constant density gradient. 
Here, the equation
(\ref{kp1})  is the evolution equation
for the nonlinear ion acoustic wave in two dimension where the unperturbed ion number density profile 
is taken to be slowly varying
and weak.

Moving into the new frame
\begin{eqnarray}
X= \xi + a(\eta), \ \
Y = \lambda, \ \
T = \eta,
\label{bendingframe}\end{eqnarray} 
with 
\begin{equation}
a(\eta) = (\frac{c_1}{1 + c_1})\int f_0(\eta) d \eta,
\label{aeta}
\end{equation}
with $c_1 = \frac{1}{\sqrt{2}}$.
 Equation (\ref{kp1}) can be transformed to the
standard constant coefficient KP equation
\begin{equation}
 \frac{\partial}{\partial X} [\frac{\partial U}{\partial \tau}
+6 U \frac{\partial U}{\partial X}
+ \frac{\partial^3 U}{\partial X^3}] +\frac{\partial^2 U}{\partial Y^2}=0,\ \
\label{kp2}
\end{equation} 
where $U= \frac{n_1}{3}$ and $\tau= T/(2+ 2 c_1)$. This is a standard completely integrable
KP equation which can be solved exactly giving  soliton solutions. But due to the presence of 
the term $a(\eta)$ which is related to $f_0(\eta)$ via (\ref{aeta}), in the new co-ordinate $X$, bending of solitons
in the two dimensional plane occurs which will be shown in the next section.

\section{Bending of solitons}

One soliton solution of KP equation (\ref{kp2}) is similar as that of
the soliton solutions of KdV equation \cite{Rec1,Rec2}
with an extra transverse direction  given by \cite{solitonkp1,solitonkp2},
\begin{equation}
 U = \frac{k_1^2}{2} Sech^2[\frac{1}{2}(k_1 X + m_1 Y - \frac{k_1^4 + m_1^2}{k_1} \tau)].
\end{equation}
Expressing the solution in old variables we get,
\begin{equation}
 n_1 = (\frac{3 k_1^2}{2}) Sech^2[\frac{1}{2}\{k_1 \xi+ k_1 a(\eta) + m_1 \lambda - \frac{k_1^4 + m_1^2}{2 k_1 (1+c_1)
} \eta\}],
\label{1soliton}
\end{equation}
with $c_1 = \frac{1}{\sqrt{2}}, $where $a(\eta)$ is given by (\ref{aeta}) and $k_1, m_1$ are arbitrary constants.

Due to the presence of the quantity $a(\eta)$ , related to the inhomogeneous ion number density , bending
of soliton occurs.

Here the plasma is inhomogeneous due to the presence of the function $f_0(\eta)$.
For different choices of $f_0$, the inhomogeneities are different.
For the choice of $f_0 = 0$ the plasma becomes homogeneous which have
the usual line solitons which is represented in FIG 2.
Thus this trivial choice of $f_0$ in the equilibrium ion number density profile
reproduces homogeneous plasma from the chosen inhomogeneity profile.

 For different functional forms  of $f_0$  dependent on the slowly varying
co-ordinate $\eta$, different types of bending occurs, which are shown in FIG. 3.

  For the sake of this 
problem, we have chosen $\delta = \epsilon$, which is the small perturbation parameter
of our calculation. The solitary wave solution is independent of
$\epsilon$, which is here taken to be $\delta$. The solution depends on the function $a(\eta)$ which is 
related to the function $f_0(\eta)$ through equation (\ref{aeta}), causing the soliton
 to bend in the two 
dimensional plane. But the amplitude of the solitary wave solution remains constant.

Similarly, the two soliton solution is given by \cite{solitonkp1,solitonkp2},
\begin{eqnarray}
n_1= 6 \frac{\partial^2}{\partial \xi^2}(\ln F_2)
\label{2soliton}, 
\end{eqnarray}
with,
\begin{eqnarray}
F_2 = 1 + e^{\eta_1} +e^{\eta_2}+ A_{12} e^{\eta_1 +\eta_2 },
 \ \ \nonumber
A_{12} = \frac{(K_1-K_2)^2 - (M_1-M_2)^2}{(K_1+K_2)^2- (M_1-M_2)^2},\\ \nonumber
\eta_1= K_1[\xi + a(\eta)+\sqrt{3} M_1 \lambda - \frac{(K_1^2 +3 M_1^2)}{(2 + 2 c_1)} \eta],\ \
\eta_2= K_2[\xi + a(\eta)+\sqrt{3} M_2 \lambda - \frac{(K_2^2 +3 M_2^2)}{(2 + 2 c_1)} \eta],
c_1= \frac{1}{\sqrt{2}}
\end{eqnarray} 
where $K_1, K_2, M_1, M_2$ are arbitrary constants. Bending of two soliton solution for different
functional forms of  $f_0(\eta)$ are also shown in FIG. 3. Since $f_0$ should reach zero value
at $\eta=0$ following FIG.1, it has been chosen accordingly.

 Now it is required to determine how much bending is taking place by
varying $f_0$ i.e, what the condition is for larger bending. 

Let us start from the one soliton solution (\ref{1soliton}). The amplitude of the 'Sech' function
is maximum when its argument goes to zero.
For static case ($\xi$ = 0), the locus of the highest amplitude of the solution is of the form
\begin{equation}
 \frac{1}{2}\{ k_1 a(\eta) + m_1 \lambda - \frac{k_1^4 + m_1^2}{2 k_1 (1+c_1)
} \eta\} = 0,
\label{locus}
\end{equation}
where $c_1 = \frac{1}{\sqrt{2}}$ .

Note that, for homogeneous plasma $f_0$ is zero making $a(\eta)$ to be also zero
determined from equation (\ref{aeta}). Hence the locus
is straight line giving line solitons for homogeneous plasma.

Now taking derivative w.r.to $\eta$ twice in the above equation (\ref{locus}) we get 
\begin{equation}
 \frac{d s}{d \eta} = -\frac{k_1 c_1}{m_1 (1+c_1)} \frac{d f_0}{d \eta}
\end{equation}
where $s = \frac{d \lambda}{\partial \eta}$ is the slope of the locus of the 
maximum amplitude. We choose $k_1,m_1$ such that
\begin{equation}
 \frac{d s}{d \eta} =  \frac{d f_0}{d \eta}
\end{equation}
 We see from the above equation that for higher value of RHS, rate of variation of slope will also be higher.
Hence the slope of the maximum amplitude curve  will vary large for traversing unit distance in $\eta$.
Larger rate of variation of slope describes larger bending.

Hence for large bending of solitons to take place, the first derivative of $f_0$ w.r.to $\eta$
must also be high.
This is incorporated in FIG 3 where bending of solitons occur for different choices of $f_0$. 

For FIG 3(a)(i), if we increase the amplitude of $f_0$ then the parabola will steepen causing larger bending.
Similar thing can be observed for FIG 3(b)(i) where the sine function becomes more rapid . Now if we increase 
the wave vector of $f_0$ in 3(b) then also the bending will become larger. Increase/decrease of both
amplitude as well as wave vector of $f_0$ will increase/decrease the first derivative of $f_0$
causing more/less bending. The same analysis can be extended to the other figures 3(c), 3(d) too.
 
\includegraphics[width=7cm]{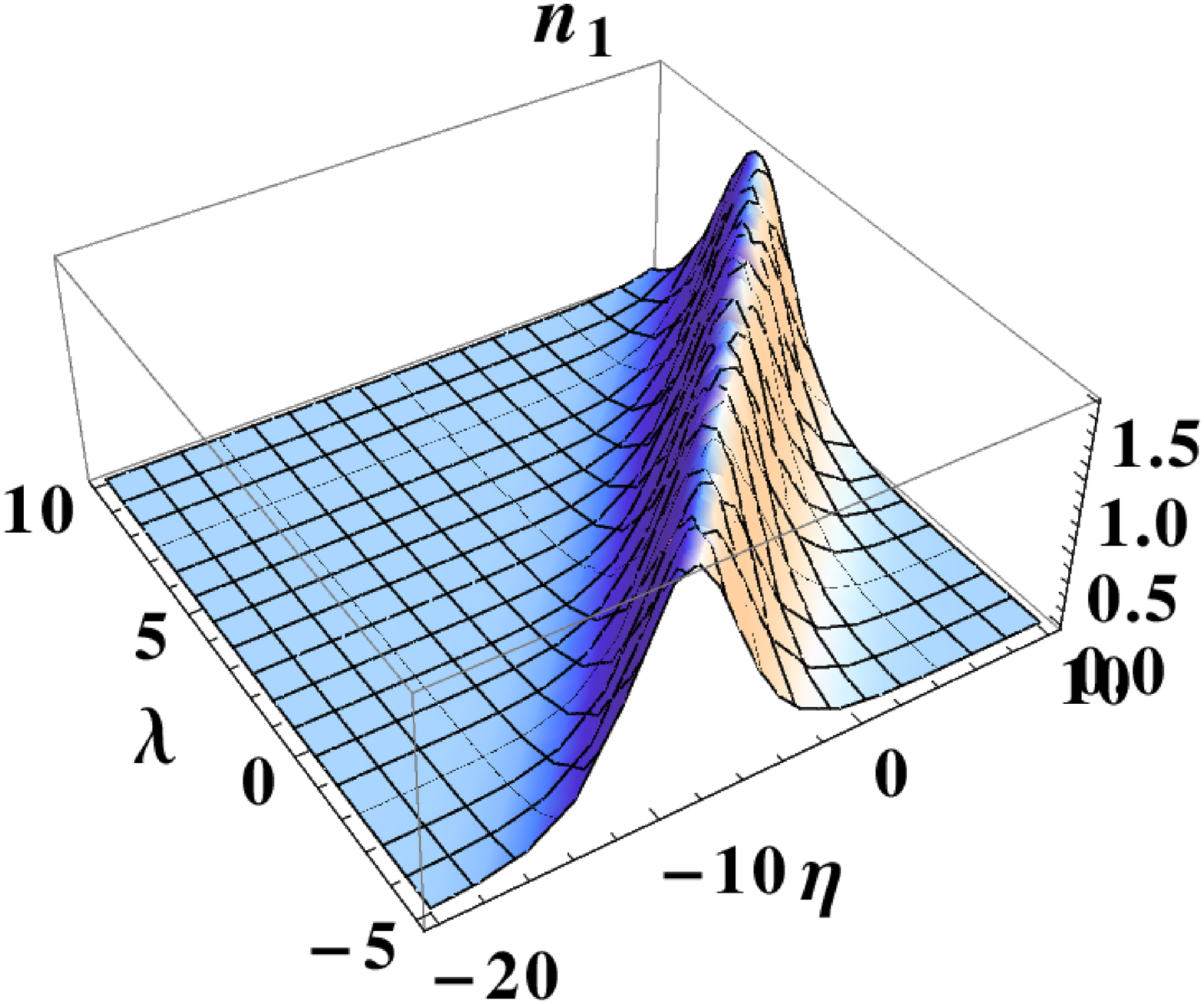}
\  \  \includegraphics[width=7cm]{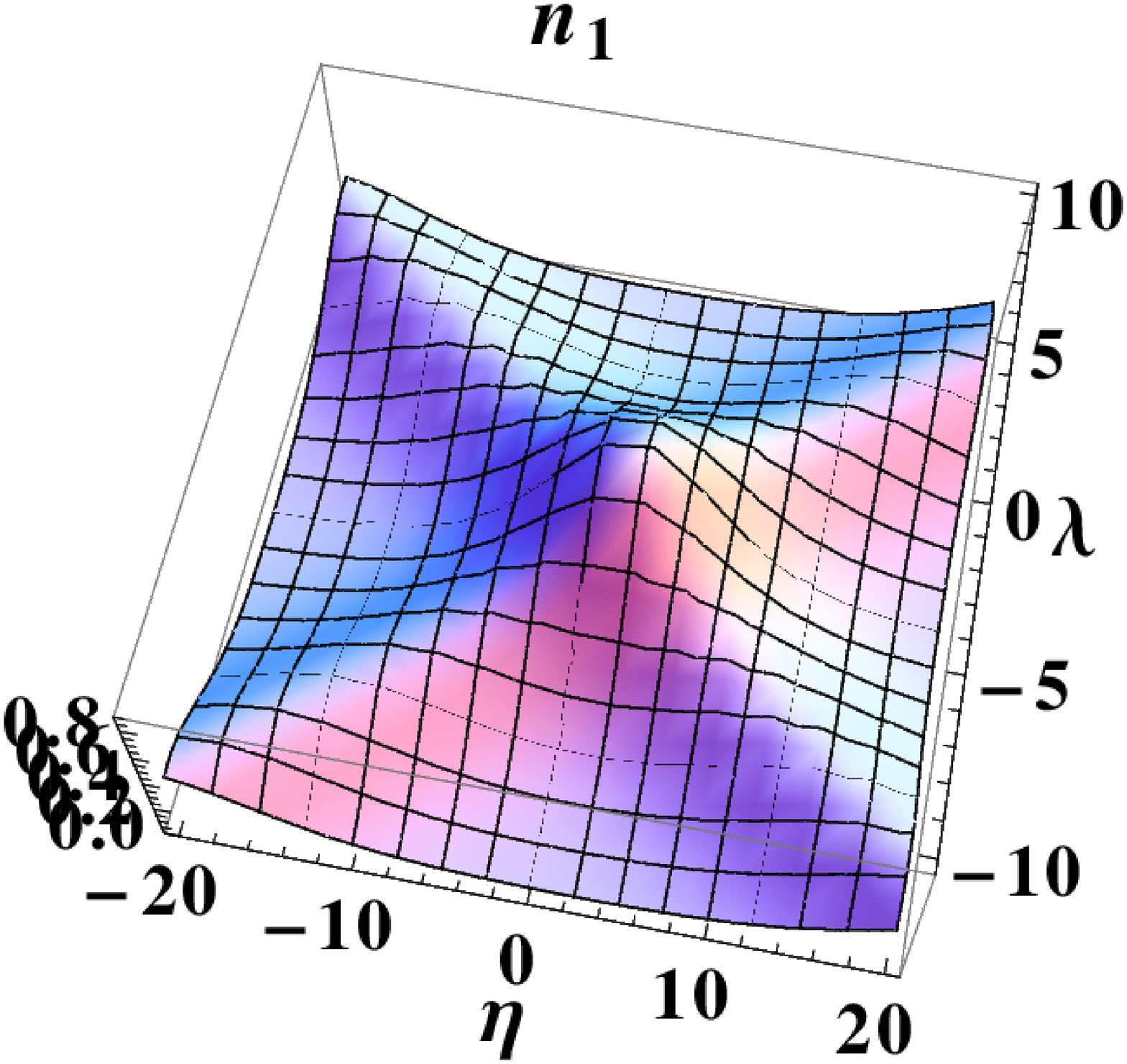}

\vspace{1cm}

\qquad \qquad (a) One soliton
\quad \qquad \qquad \quad \qquad \qquad \qquad \qquad \qquad(b) Two soliton
 
\vspace{1cm}

\noindent FIG.2: {\bf Static piture of  one and two soliton solutions given by
(\ref{1soliton}) and (\ref{2soliton}) of the two dimensional ion acoustic wave 
at $\xi=0$ for $k_1 = 1, m_1 = 1, K_1 = K_2 = \frac{1}{2},
P_1 = -P_2 = \frac{2}{3}$ and $f_0 =0 $. Clearly, $f_0 = 0$ represents homogeneous plasma
which is reproduced here from the chosen inhomogeneous ion number density profile. Line solitons
which are found in the homogeneous plasma are also reproduced here for this trivial choice of $f_0$. }

\vspace{1cm}

\includegraphics[width=7cm]{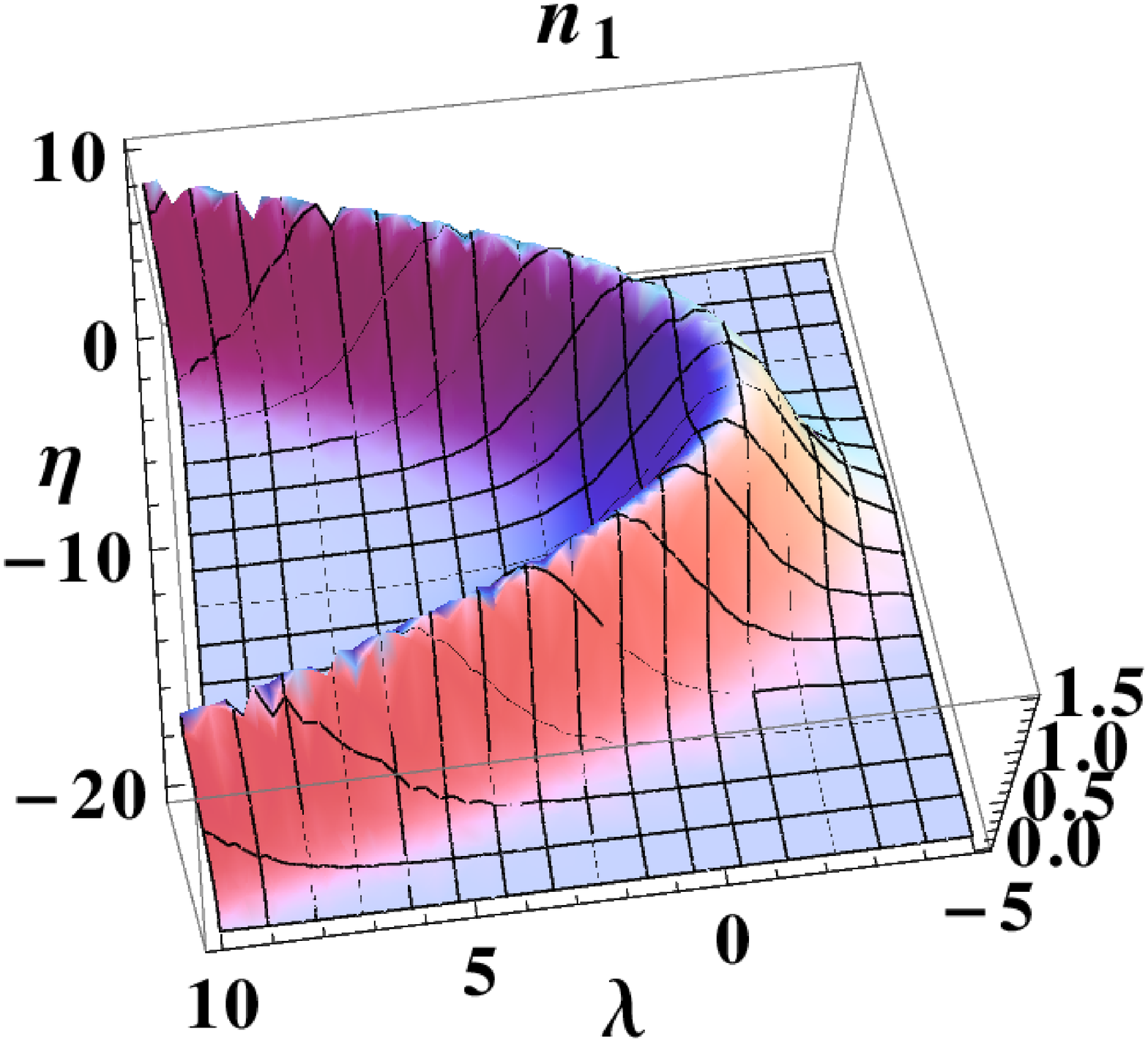}
 \ \ \ \includegraphics[width=7cm]{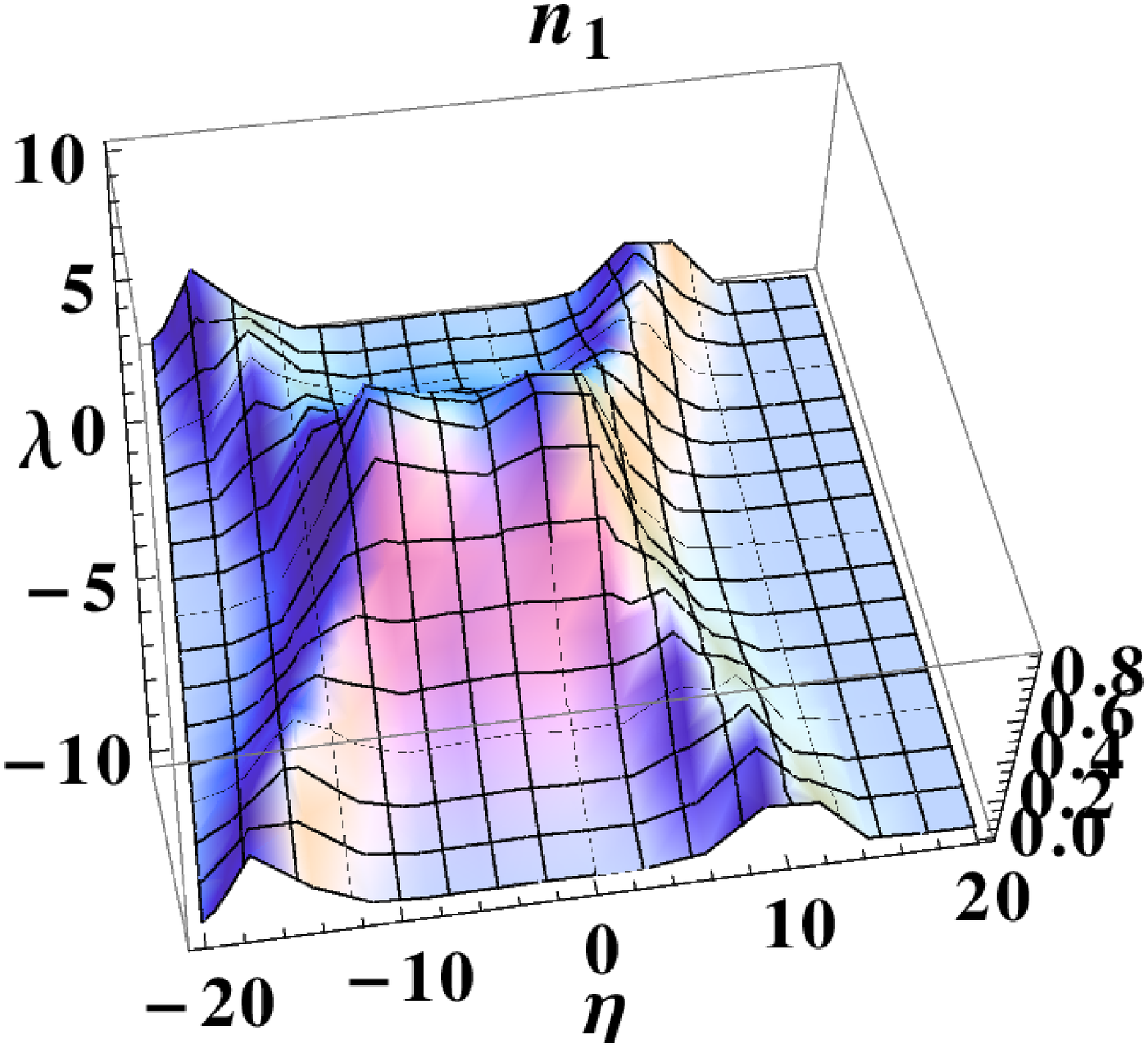}

\vspace{1cm}

\qquad \qquad (a)(i) One soliton for $ f_0 = -\eta/3$
\quad \qquad \qquad \quad \qquad (a)(ii)Two soliton 
for $ f_0 = -\eta/3$

\vspace{1cm}

\includegraphics[width=7cm]{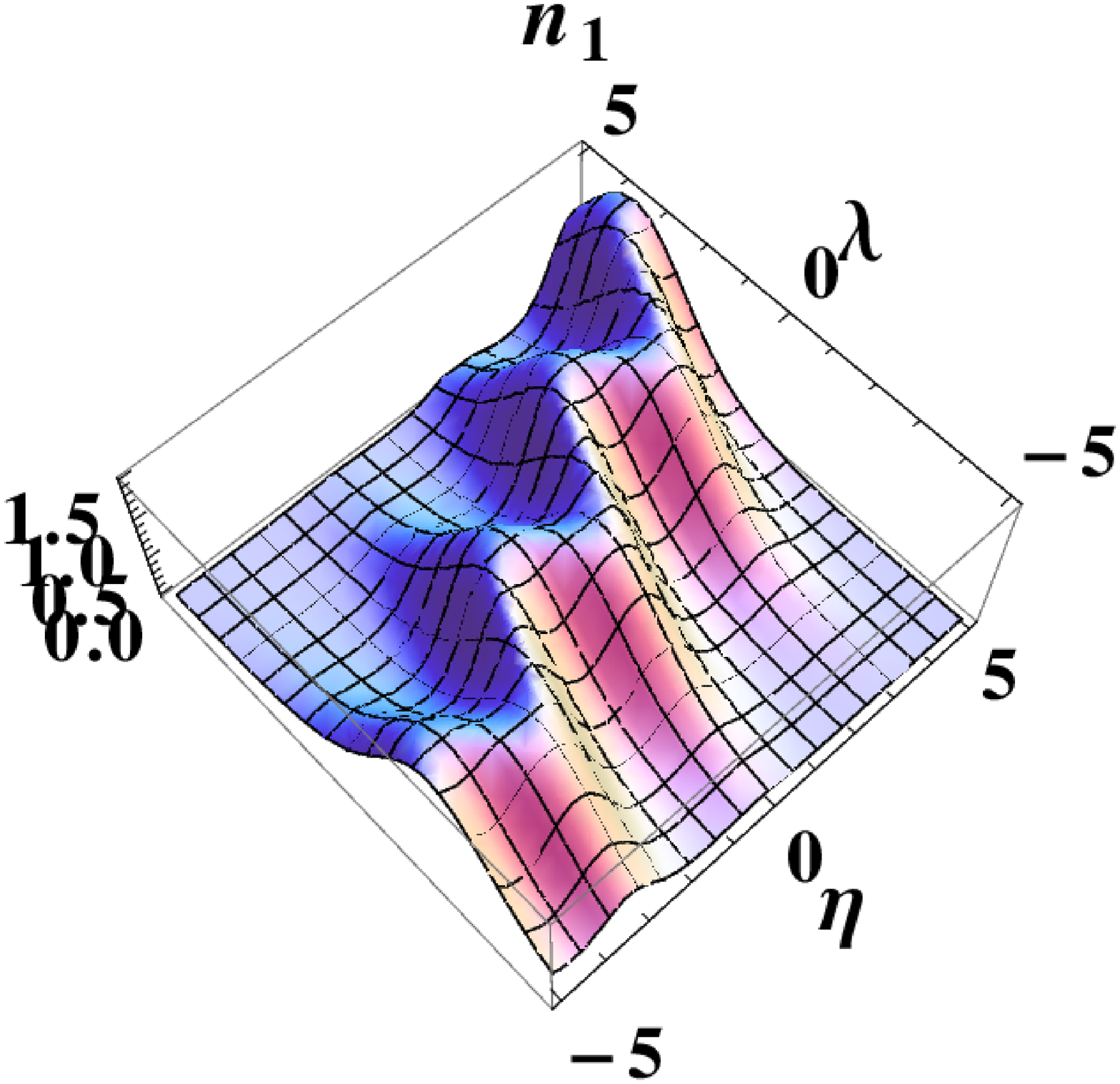}
 \ \ \ \includegraphics[width=7cm]{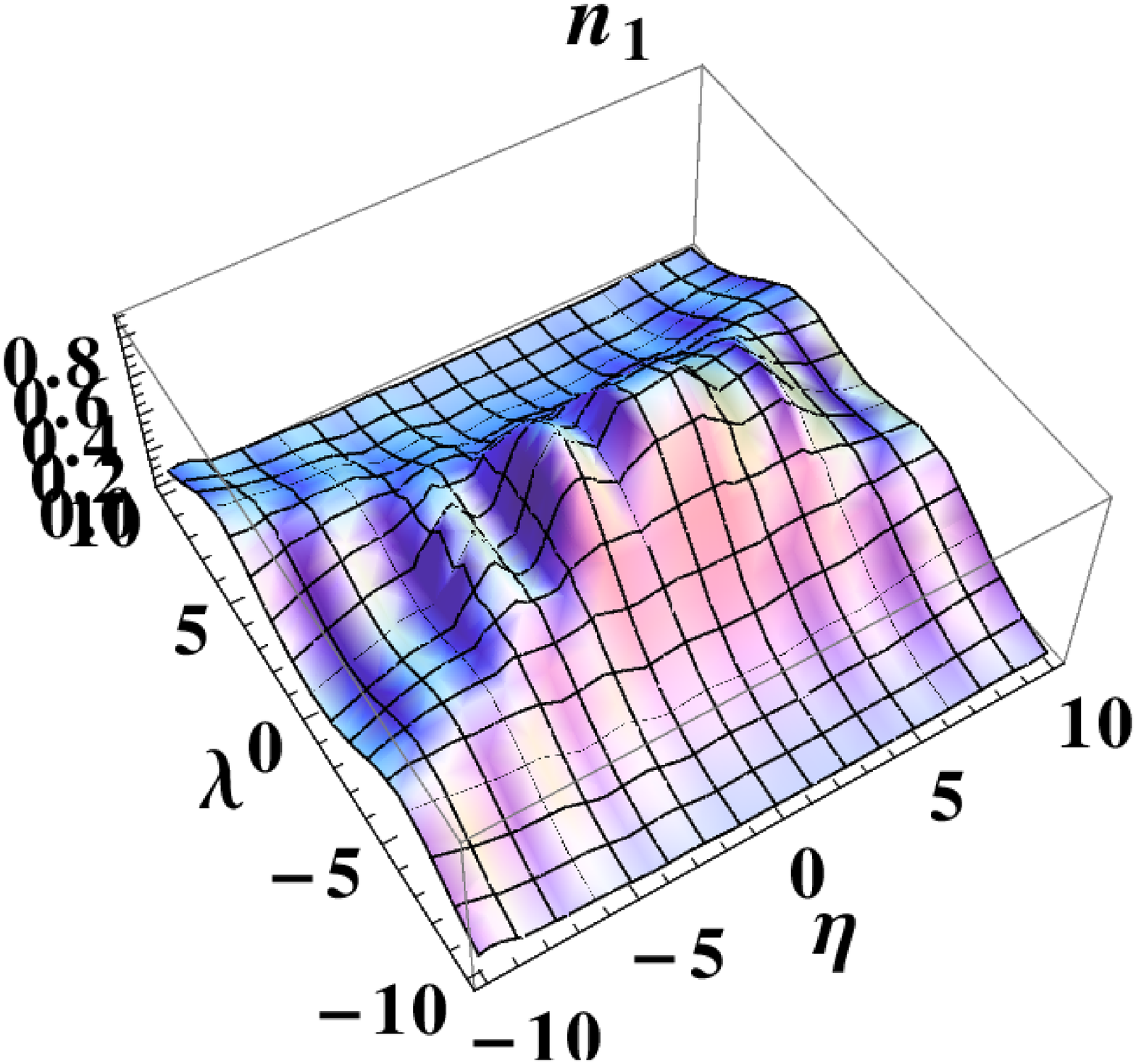}

\vspace{1cm}

 (b)(i) One soliton for $ f_0 = - 4 Sin(2 \eta)$
\quad \qquad \qquad  (b)(ii)Two soliton for $ f_0 = - 4 Sin(2 \eta)$

\vspace{1cm}

\includegraphics[width=7cm]{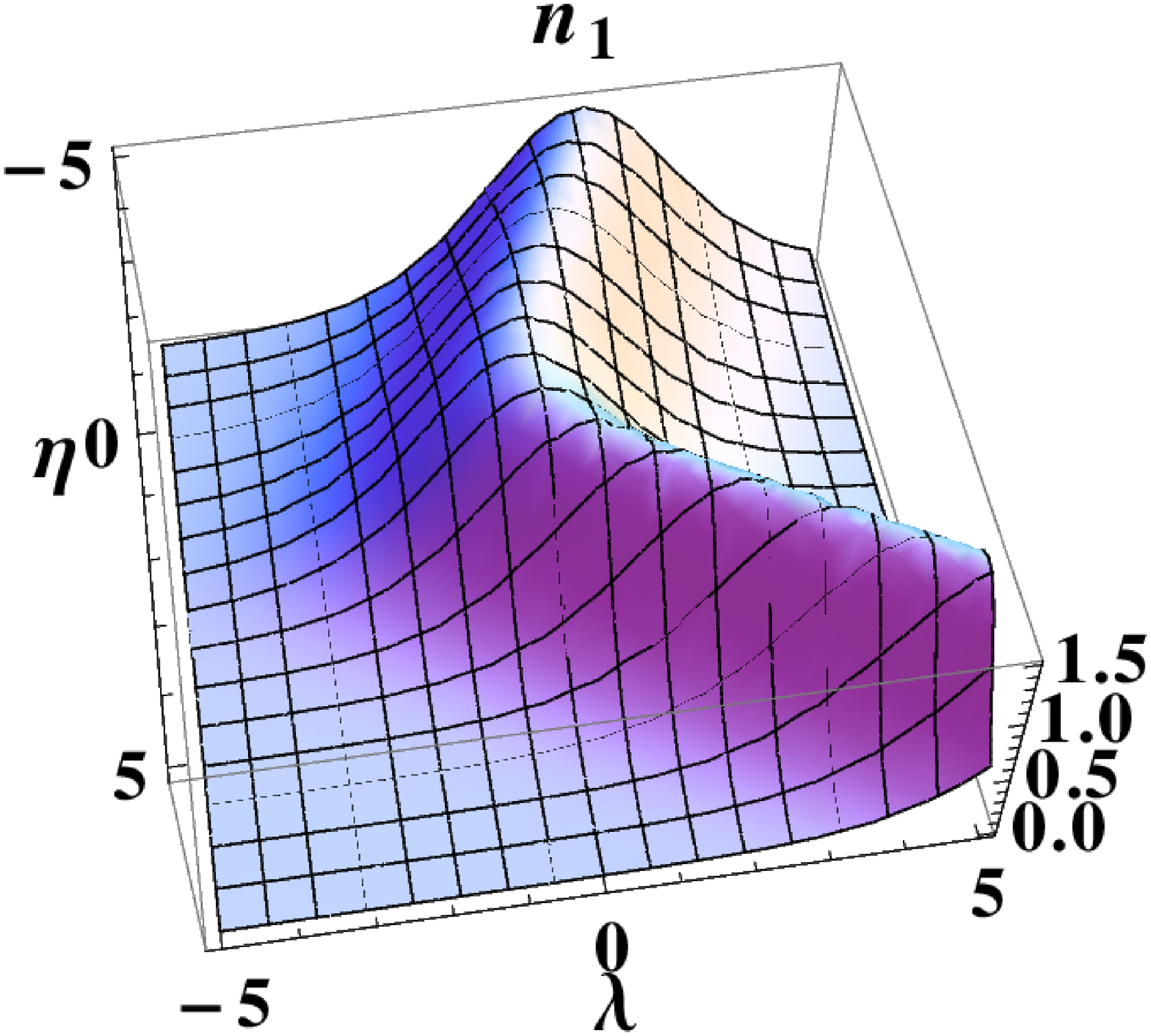}
 \ \ \ \includegraphics[width=7cm]{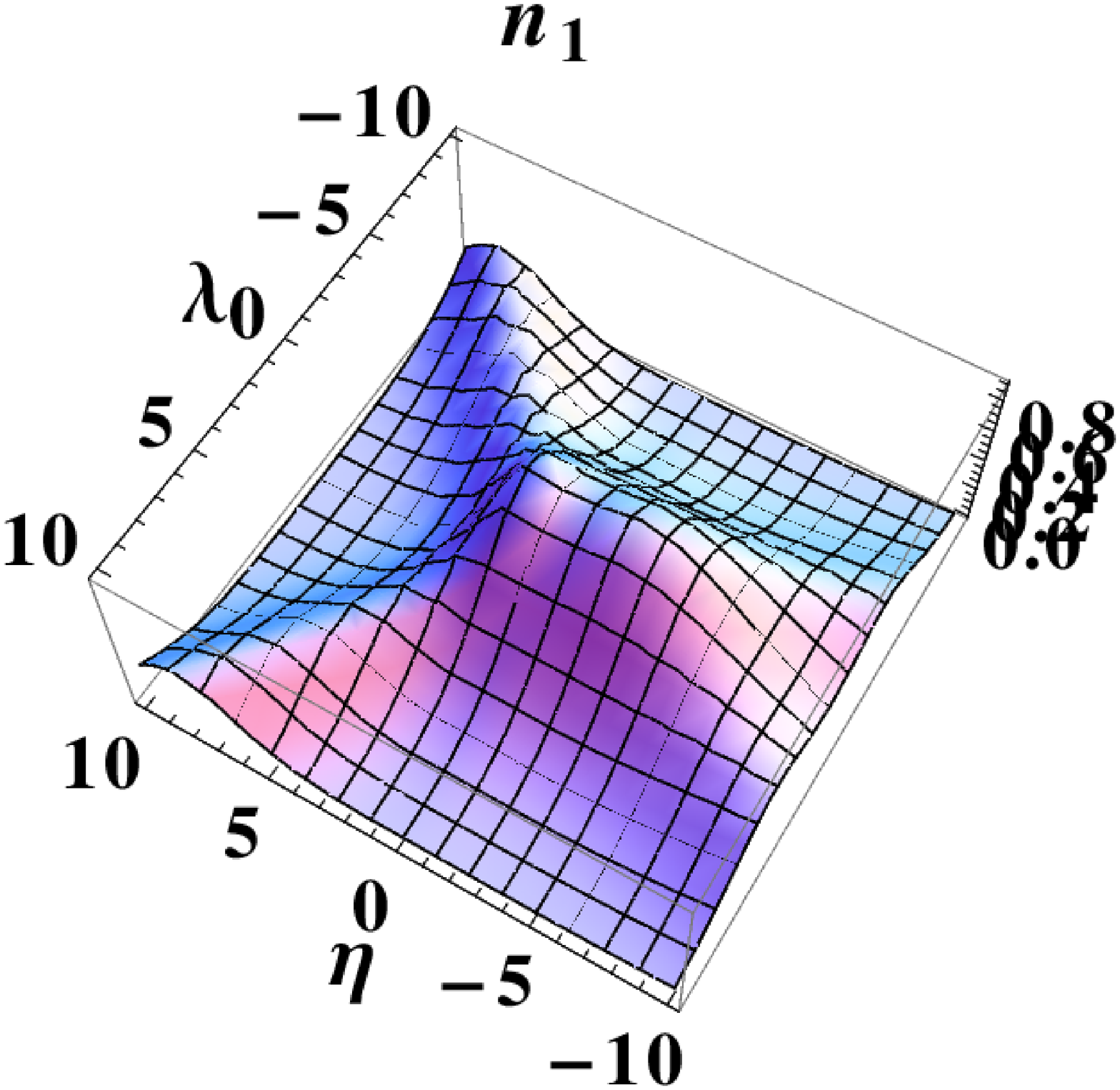}

\vspace{1cm}

 (c)(i) One soliton for $ f_0 = - (1+\sqrt{2}) tanh(\eta)$
\quad \qquad \qquad  (c)(ii)Two soliton for $ f_0 = - (1+\sqrt{2}) tanh(\eta)$

\vspace{1cm}

\includegraphics[width=7cm]{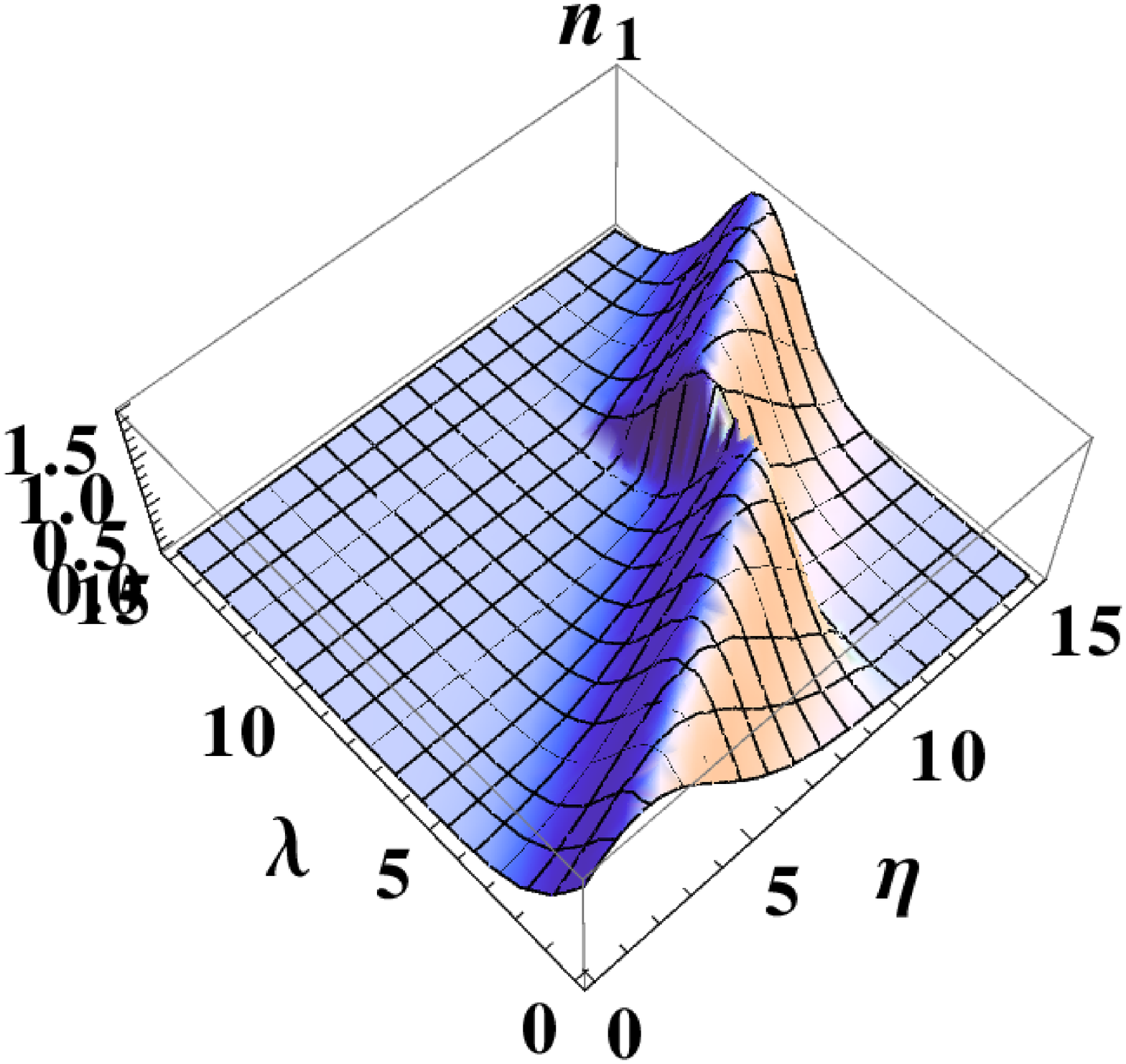}
 \ \ \ \includegraphics[width=7cm]{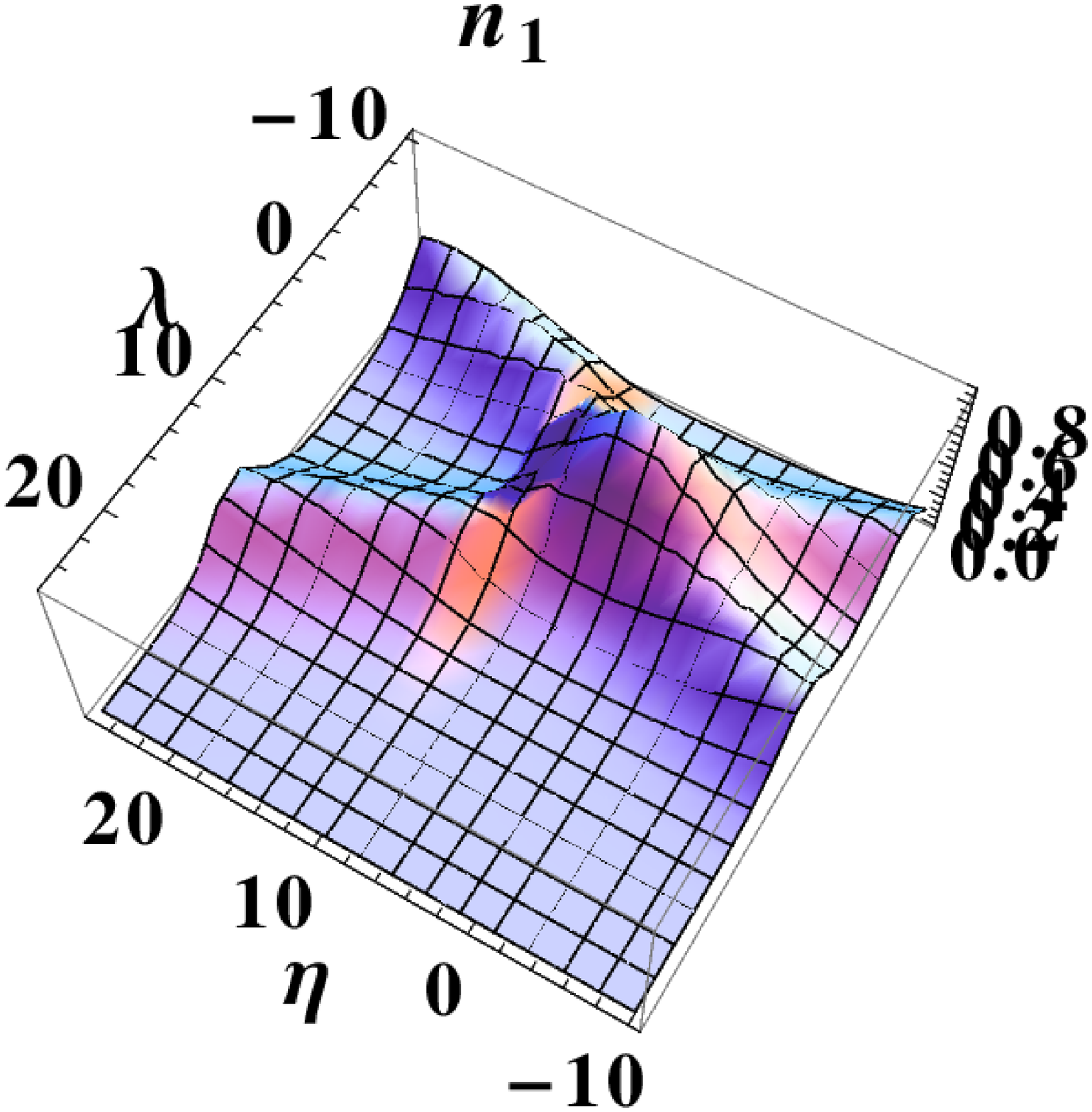}

\vspace{1cm}

 (d)(i) One soliton 
  \quad \qquad \qquad \quad \qquad \qquad \quad \qquad \qquad \quad \qquad \qquad (d)(ii)Two soliton 

for $ f_0 = -5 (\sqrt{2}+1) (Sech(5(\eta-10)))^2$)
  \quad \qquad \qquad  \quad \qquad \qquad for  $ f_0 = - -5 (\sqrt{2}+1) (Sech(5(\eta-10)))^2$)
\vspace{1cm}

\noindent FIG.3: {\bf Static piture of  one and two soliton solutions given by
(\ref{1soliton}) and (\ref{2soliton}) of the two dimensional ion acoustic wave 
at $\xi=0$ for $k_1 = 1, m_1 = 1, K_1 = K_2 = \frac{1}{2},
P_1 = -P_2 = \frac{2}{3}$ and for the specified functions of  $f_0$
which is related to unperturbed ion number density.
 The different functional forms of $f_0$ causes the phase of the solitary wave to change which causes bending
in the two dimensional plane, whereas the amplitude remains constant.}

\vspace{1cm}  
Frycz and Infeld obtained the bending of soliton \cite{Bending1} by studying numerically
the nonlinear stability analysis of KP equation. The characteristics of KP equation state that the
 initial condition must fulfill
an infinite set of constraints if the solution is to remain localized. Just adding a
 perturbation to one soliton solution would violate this constraint. Thus bending is a natural
perturbation which is a choice for initial condition of this numerical simulation.
But in our work, the bending of solitons were obtained  analytically showing dependence on $f_0$
which is related to inhomogeneous ion number density.
We have exactly solved the KP equation (\ref{kp1}) obtained for the two dimensional 
propagation of ion acoustic wave for  weak and slowly varying
 inhomogeneity, related to the arbitrary function
$f_0(\eta)$. Since we have transformed the evolution equation into a standard constant coefficient KP equation,
its each and every solution faces the same phase modification controlled by $f_0$,
causing the shape of the solution to change in the two dimensional plane. 
The amplitude of the soliton solutions is found to  remain constant. This is in view of the weak and slowly varying
inhomogeneous ion number density, so that all variations appear only in the phase of the soliton.

 We see that
the weak and slowly varying equilibrium potential, which exists in the plasma, is a function of $f_0$,
varying along the x axis (i.e, $\eta$ axis). Hence due to this time independent potential
an electric field develops which exerts force on the
ions, constituting the soliton. But due to the inhomogeneity of the equilibrium potential
function, different ions situated at different positions are attracted (or repelled) differently.
Again an equilibrium ion drift velocity also exists, which is also directed in the 
x axis and inhomogeneous. Due to the superposed effects of the inhomogeneous equilibrium and also the 
time dependent quantities, the  ions change their positions. This causes the ion 
acoustic soliton to bend in the two dimensional plane. Since the potential drop is weak as well as 
slow, the number of ions forming soliton do not change drastically. Hence the amplitude of the soliton remains 
constant causing its  phase to vary with $f_0(\eta)$.

We see that, the one soliton solution  of our evolution equation (\ref{kp1})
contains the inhomogeneous function $a(\eta)$ in the phase. We see from the solution that
the function reaches its maximum value when the phase factor turns to be zero. Hence as the soliton
propagates in the two dimensional plane, $\eta$ changes causing $a(\eta)$ to change nonlinearly 
depending on $f_0(\eta)$. Now if we fix the time variable $\xi$, then the transverse variable $\lambda$ 
has to adjust itself in order to make the phase factor of the ''Sech``
function zero causing soliton bending. 

%

 These bending features of the solitons is very relevant and important
in the context of inhomogeneous plasmas along with the other features like reflection, transmission etc. But
such a feature  has not been explored till now. We see in this work that if the equilibrium density variation
is slow and weak,  which is very close to the homogeneous value then these bending features
can be seen. Hence a more accurate experiment could reveal such finer effects.

\section{Conclusive remarks}
In this work,  we have obtained the bending of ion acoustic  solitary wave in the two dimensional
plane for the propagation in unmagnetized plasma with cold ions and isothermal electrons with weak and
slowly varying density inhomogeneity. We have obtained a modified KP equation with an extra term arising
due to inhomogeneous equilibrium ion density . We have exactly solved the KP equation 
giving a solitary wave solution in which the phase of the soliton
gets modified by a function $f_0$, which is related to unperturbed ion density,
 causing soliton bending, where as the amplitude remains constant.
The bending features of the solitons is very relevant and important
in inhomogeneous plasma along with the other features like reflection, transmission etc. 
 More accurate and precession experiment could reveal such finer and interesting features.


\begin{thebibliography}{99}
\bibitem{Washimi}
H. Washimi and T. Taniuti, Phys. Rev. Lett  17, 19 (1966). 

\bibitem{Chen}
F.F Chen
 {\it Introduction to Plasma physics and controlled fusion },
(Plenum Press, New York and London, 1984).
\bibitem{Farah} 
 F. Aziz
{\it Thesis: Ion-acoustic solitons: Analytical, experimental and numerical studies},  (2011) and references therein
\bibitem{Kaw}
N. Nishikawa and  and P. K.Kaw Phys. Lett  A 50, 455 (1975). 
\bibitem{Gell} Y. Gell and L. Gomberoff, 
Phys. Lett. A 60, 125 (1977).
\bibitem{JohnSaxena}
 P.I John and Y.C Saxena Phys. Lett. A  56, 385 (1976).
 \bibitem{RaoVerma} N. N Rao and R. K Verma, 
.Phys. Lett. A  70, 9 (1979).
 \bibitem{KuehlImen}
H.H Kuehl and K Imen,  Phys. Fluids.   28, 2375 (1985).
\bibitem{Chang}
H.Y Chang, S. Raychaudhuri, J.Hill, E.K Tsikis and K.E Lonngren,  Phys. Fluids.   29, 294 (1986).
\bibitem{Dahiya2}
R. P Dahiya, P.I John and Y. C Saxena,  Phys. Lett. A  65, 323 (1978).

\bibitem{Popa}
G. Popa and M.Oertl, Phys. Lett. A  98, 110 (1983).
\bibitem{Nishida}
Y. Nishida,  Phys. Fluids  27, 2176 (1984).
\bibitem{ImenKuehl}
K.Imen and H.H Kuehl,  Phys. Fluids  30, 73 (1987).  
\bibitem{Nagasawa} T. Nagasawa and Y. Nishida, 
Phys. Rev. Lett 56, 2688 (1986).
\bibitem{Seungjun}
S. Yi, J.L Cooney, H.Kim, A. Amin, Y. El-Zein and K.E Lonngren, Phys. Plasmas  3, 529 (1996).
\bibitem{Cooney}
J.L Cooney, M.T Gavin, J.E Williams, D. W Aossey and K.E Lonngren,  Phys. Fluids B  3, 3277 (1991).
\bibitem{Kuehl} H.H Kuehl
Phys. Fluids 26, 1577 (1983).
\bibitem{KoKuehl}
K.Ko and H.H Kuehl,  Phys. Rev. Lett  40, 233 (1978).
\bibitem{KoKuehl2}
 K. Ko and H.H Kuehl, Phys. Pluids  23, 834 (1980).
\bibitem{SinghDahiya} S.Singh and R.P Dahiya,  
Phys. Fluids B 3, 255 (1991).
\bibitem{Singhdahiya2}
S.Singh and R.P Dahiya,  
J.Plasma Phys 41, 185 (1989).
\bibitem{Malik1}
H.K Malik and R.P Dahiya, Phys.Plasma 1, 2872(1994)
\bibitem{Malik6}
D.K Singh and H.K Malik , Phys.Plasma 13, 082104(2006).
\bibitem{Chauhan} 
S.S Chauhan,H.K. Malik and R.P. Dahiya, 
Phys. Plasma 3, 3932 (1996).
\bibitem{SinghMalik} D.K Singh and H.K Malik,  Phys. Plasma 14, 062113 (2007).
\bibitem{Xiao}
D.Xiao, J.X Ma, Y.Li, Y.Xia and M.Y Yu, Phys. plasma 13,052308 (2006)
\bibitem{Kumar}
 R.Kumar, H.K Malik, and S.Kawata, Physica. D 240,310 (2011)
\bibitem{Malik3} H.K Malik , 
 Phys. Lett. A 365, 224 (2007).
\bibitem{Mahmood} 
Q.Haque and S.Mahmood, Phys. Plasmas 15,034501 (2008)
\bibitem{Malik4} 
H.K Malik, Phys. plasma 15,072105 (2008)
\bibitem{Malik5}
Jyoti and H.K Malik Phys. Plasmas  18, 102116 (2011).
\bibitem{Malik8} H.K Malik, Jyoti and R.Kumar, 
J.Theor. Appl Phys 8, 123 (2014).
\bibitem{Malik7}
K.Singh, V.Kumar and H.K Malik  Phys. Plasmas  12, 072302 (2005).
\bibitem{Duan-Zhao}
 W.Duan and J.Zhao  Phys. Plasmas  6, 3484 (1999).
\bibitem{IbrahimKuehl}
I. Ibrahim and H.H Kuehl, Phys.Fluids 27,962 (1984)
\bibitem{ZK2} 
W.F El-Taibany,M.M Selim, N.A El-Bedwehy and O.M Al-Abbasy,  Phys. Plasmas  21, 073710 (2014).
\bibitem{ZK3}
A.P Misra and A.R Chowdhury  Phys. Plasmas  13, 062307 (2006).
\bibitem{ZK4}
W.Masood  Phys. Plasmas  17, 052312 (2010).
\bibitem{ZK1}
W.F.El-Taibany, M.Wadati and R. Sabry , Phys. Plasmas  14, 032304 (2007).
\bibitem{KP}
B.B Kadomtsev and V.I Petviashvili,  Sov. Phys.Dokl.15, 539 (1970).
\bibitem{KP2}
 H.K Malik, S.Singh and R.P dahiya,  Phys. Lett.A  195, 369 (1994).
\bibitem{KP3}
W. Masood, Phys. Lett.A  373, 1455 (2009).
\bibitem{YANG} 
J.R Yang, X.Y Tang, X.N Gao, X.P Cheng and S.Y. Lou
EPL 12, 45001 (2011).
\bibitem{ZhangXue} L.P. Zhang and J.K. Xue  
Commun Nonlinear Sci Numer Simulat 15, 3379 (2010).
\bibitem{Bending1} P. Frycz and E.Infeld ,  Phys. Rev. A  41,3375 (1990).
\bibitem{solitonkp1}
A.M Wazwaz  Appl.Math.Comput,  190, 633 (2007).
\bibitem{solitonkp2}
 M.J Ablowitz and D.E Baldwin, Phys.Rev.E  86, 036305 (2012).
\bibitem{Rec1}
H.Saleem, S.Ali and Q.Haque , Phys. Plasmas  22, 084509 (2015).
\bibitem{Rec2}
R.Jahangir, W.Masood, M.Siddiq, N.Batool and K.Saleem , Phys. Plasmas  22, 092312 (2015).
\end{thebibliography}
\end{document}